\documentclass{aa}  

\usepackage{graphicx}
\usepackage{float}
\usepackage{txfonts}
\usepackage[switch]{lineno}

\begin{document}

\title{Hadronic particle acceleration in the supernova remnant SN 1006 as traced by \emph{Fermi}-LAT observations}

\titlerunning{Hadrons in SN 1006 as traced by \emph{Fermi}-LAT observations}

   \author{M. Lemoine-Goumard\inst{1} \and 
           F. Acero\inst{2, 3}
          \and
          J. Ballet\inst{2} \and
          M. Miceli\inst{4, 5}
          }

   \institute{Univ.  Bordeaux, CNRS, LP2i Bordeaux, UMR 5797, F-33170 Gradignan, France\\
    \email{lemoine@lp2ib.in2p3.fr}
   \and 
   AIM, CEA, CNRS, Universit\'e Paris-Saclay, Universit\'e de Paris, F-91191 Gif sur Yvette, France \\
          \and
     FSLAC IRL 2009, CNRS/IAC, La Laguna, Tenerife, Spain  \\     
         \and
    Dipartimento di Fisica e Chimica E. Segr\`e, Universit\`a degli Studi di Palermo, Piazza del Parlamento 1, 90134, Palermo, Italy\\
    \and
    INAF-Osservatorio Astronomico di Palermo, Piazza del Parlamento 1, 90134, Palermo, Italy
             }

   \date{}

  \abstract{The supernova remnant SN 1006 is a source of high-energy particles detected at radio, X-rays, and tera-electronvolt gamma rays. It was also announced as a source of gamma rays by \emph{Fermi}-LAT but only the north-east (NE) limb was detected at more than 5$\sigma$ significance level. Using 15 years of \emph{Fermi}-LAT observation and a thorough morphological analysis above 1 GeV, we report the detection of the NE rim at the 6$\sigma$ level and the south-west (SW) rim at the 5.5$\sigma$ level using radio templates from the GLEAM survey. The spectral analysis performed between 100 MeV and 1 TeV allows the detection of a hard spectral index for the NE limb of $1.7 \pm 0.1 \pm 0.1$ while the emission detected in the SW is well reproduced with a steeper spectral index of $2.2 \pm 0.1 \pm 0.1$. A marginal detection ($\sim 3\sigma$) of emission coincident with the bright north-west (NW) H$\alpha$ filament is also described with a similar spectral index of $\sim 2.1$.
  We successfully characterized the non-thermal multi-wavelength emission of the NE and SW limbs with a model in which inverse-Compton emission dominates in the NE while proton-proton interactions becomes significant in the SW due to the enhanced density of the medium. 
  }


   \maketitle
%
\begin{figure}[ht!]
\includegraphics[width=\linewidth]{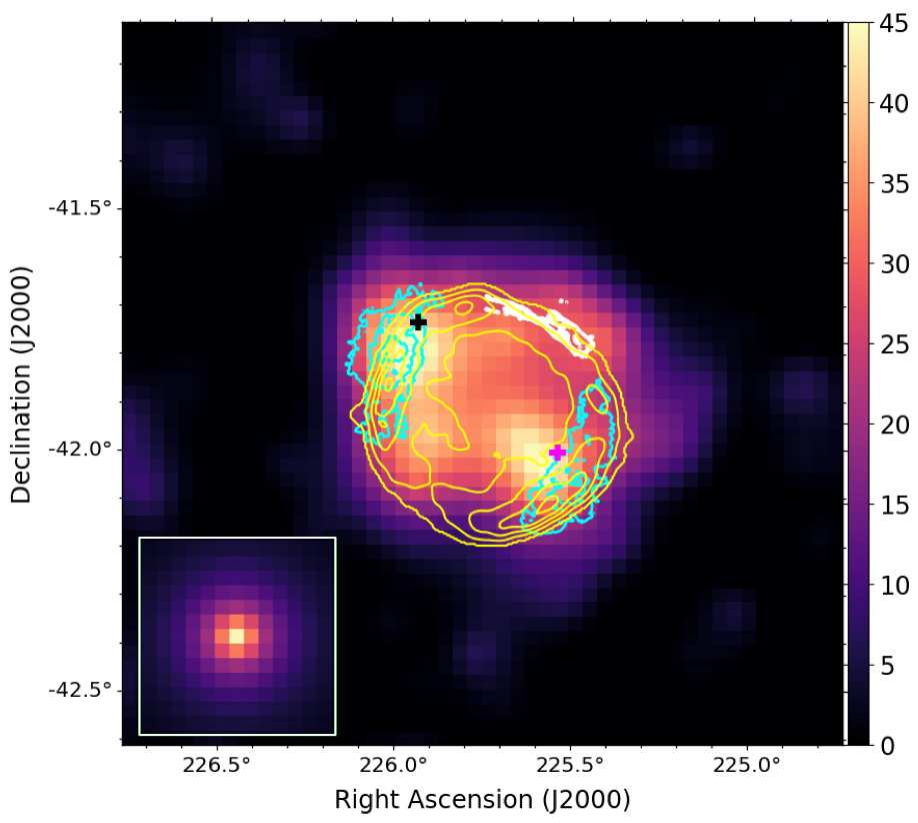}
\caption{\emph{Fermi}-LAT TS map above 1 GeV for the $1.5 \degr \times 1.5 \degr$ region around the SNR SN 1006. The black cross indicates the only 4FGL-DR4 source in the region (4FGL J1503.6$-$4146) while the magenta cross indicates the point source coincident with the SW rim added in our best model. The cyan contours represent the H.E.S.S. significance contours at 3, 5, 7$\sigma$ \citep{2010A&A...516A..62A}. White contours show the H$\alpha$ template generated from the 4m Blanco telescope observations at CTIO \citep{2014ApJ...781...65W}. Yellow contours present the radio spatial template derived using observations from the Murchison Widefield Array \citep{2017MNRAS.464.1146H}. The inset on the bottom left corner provides the counts map from a simulated point source in the same coordinate system (more details in Appendix~\ref{sec:ap}).}
\label{fig:tsmap1}
\end{figure}

\section{Introduction} \label{sec:intro}
Located at 2.2 kpc from Earth \citep{2003ApJ...585..324W}, the Type Ia supernova remnant (SNR) SN 1006 is one of the few historical remnants observed from Earth. It is an ideal target to study the Fermi acceleration process in astrophysical shocks. Indeed, it was the first SNR in which a non-thermal component of hard X-rays was detected in the rims of the remnant by \emph{ASCA} \citep{1995Natur.378..255K}, providing a clear evidence for diffusive shock acceleration of electrons to high energies in the north-east (NE) and south-west (SW) limbs. Indications of efficient hadronic acceleration in the nonthermal limbs have been also provided recently \citep{2022NatCo..13.5098G}. High resolution images by \emph{Chandra} then revealed small-scale structure in the non-thermal X-ray filaments of the NE rim of SN 1006 \citep{2003ApJ...586.1162L}, supporting the idea of high magnetic fields in the bright limbs of the remnant. Deep observations at very high energy (VHE: above 100 GeV) were carried out with the H.E.S.S. telescopes from 2003 to 2008 allowing the detection of a bipolar morphology, strongly correlated with the non-thermal X-rays \citep{2010A&A...516A..62A}. The two H.E.S.S. sources J1504$-$418 and J1502$-$421 correspond to the NE and SW shell regions and share similar flux values. Using 3.5 and 6 years of \emph{Fermi}-Large Area Telescope (LAT) data, only upper limits have been obtained by \cite{Araya2012} and \cite{2015A&A...580A..74A}, respectively. The detection at $4\sigma$ level of a $\gamma$-ray source coincident with SN 1006  was claimed by \cite{Xing2016} using 7 years of LAT data. This was then confirmed by \cite{Condon_2017} using 8 years of data which allowed the detection of the SNR at $6\sigma$ as well as indication of an asymmetry of the high-energy $\gamma$-ray emission between the NE and SW regions. \cite{2019PASJ...71...77X} finally announced the detection of the SW limb at a $4\sigma$ significance level using 10 years of data. By performing a broadband SED modeling of the two limbs, the authors concluded that, similarly to the case of the NE limb, the gamma-ray emission from the SW limb is likely dominated by the leptonic process in which high-energy electrons accelerated from the shell of the SNR inverse-Compton scatter background photons.\\
In this work we aim to characterize the $\gamma$-ray emission detected at giga-electronvolt energies with \emph{Fermi}-LAT using 15 years of observations (Section~\ref{sec:obs}), undertake a full morphological analysis of the SNR (Section~\ref{sec:morpho}) and a spectral analysis of the different spatial components detected (Section~\ref{sec:spec}). The results are then discussed using a one-zone modeling of the multi-wavelength data (Section~\ref{sec:model}).

\section{\emph{Fermi}-LAT observations} \label{sec:obs}
The \emph{Fermi}-LAT is a $\gamma$-ray telescope that detects photons by converting them into electron-positron pairs in the range from 20 MeV to higher than 500 GeV \citep{2009ApJ...697.1071A}. The following analysis is performed using 15 years of \emph{Fermi}-LAT data (2008 August 04 -- 2023 August 03) centered on SN 1006. Time intervals during which the satellite passed through the South Atlantic Anomaly are excluded. Our data are also filtered removing time intervals around solar flares and bright GRBs, following the procedure used in all \emph{Fermi}-LAT catalogs. The current version of the LAT data is P8R3 \citep{bruel2018fermilatimprovedpass8event}. We use the SOURCE class event selection, with the instrument response functions P8R3\_SOURCE\_V3. The Galactic diffuse emission is modeled by the standard file gll\_iem\_v07.fits and the residual background and extragalactic radiation are described by a single isotropic component with the spectral shape in the tabulated model iso\_P8R3\_SOURCE\_V3\_PSFn\_v1.txt. The models are available from the \emph{Fermi} Science Support Center (FSSC)\footnote{\url{https://fermi.gsfc.nasa.gov/ssc/data/access/lat/BackgroundModels.html}}.
The data reduction and exposure calculations are performed using the LAT $fermitools$ version 2.2.0 and $fermipy$ \citep{2017ICRC...35..824W} version 1.2.0. We perform a binned likelihood analysis with 10 energy bins per decade over a region of $15^{\circ} \times 15\fdg$ We included all sources from the LAT 14-year source Catalog (4FGL-DR4)\footnote{\url{https://fermi.gsfc.nasa.gov/ssc/data/access/lat/14yr_catalog}} in a region of $25^{\circ} \times 25\fdg$ We account for the effect of energy dispersion (when the reconstructed energy differs from the true energy) by setting the parameter ${\rm edisp\_bins}=-2$. As such, the energy dispersion correction operates on the spectra with two extra bins below and above the threshold of the analysis\footnote{The energy dispersion correction is applied to all sources in the model, except for the isotropic diffuse emission model. More details can be found in the FSSC: \url{https://fermi.gsfc.nasa.gov/ssc/data/analysis/documentation/Pass8_edisp_usage.html}}. 

\begin{figure}[ht]
\includegraphics[width=\linewidth]{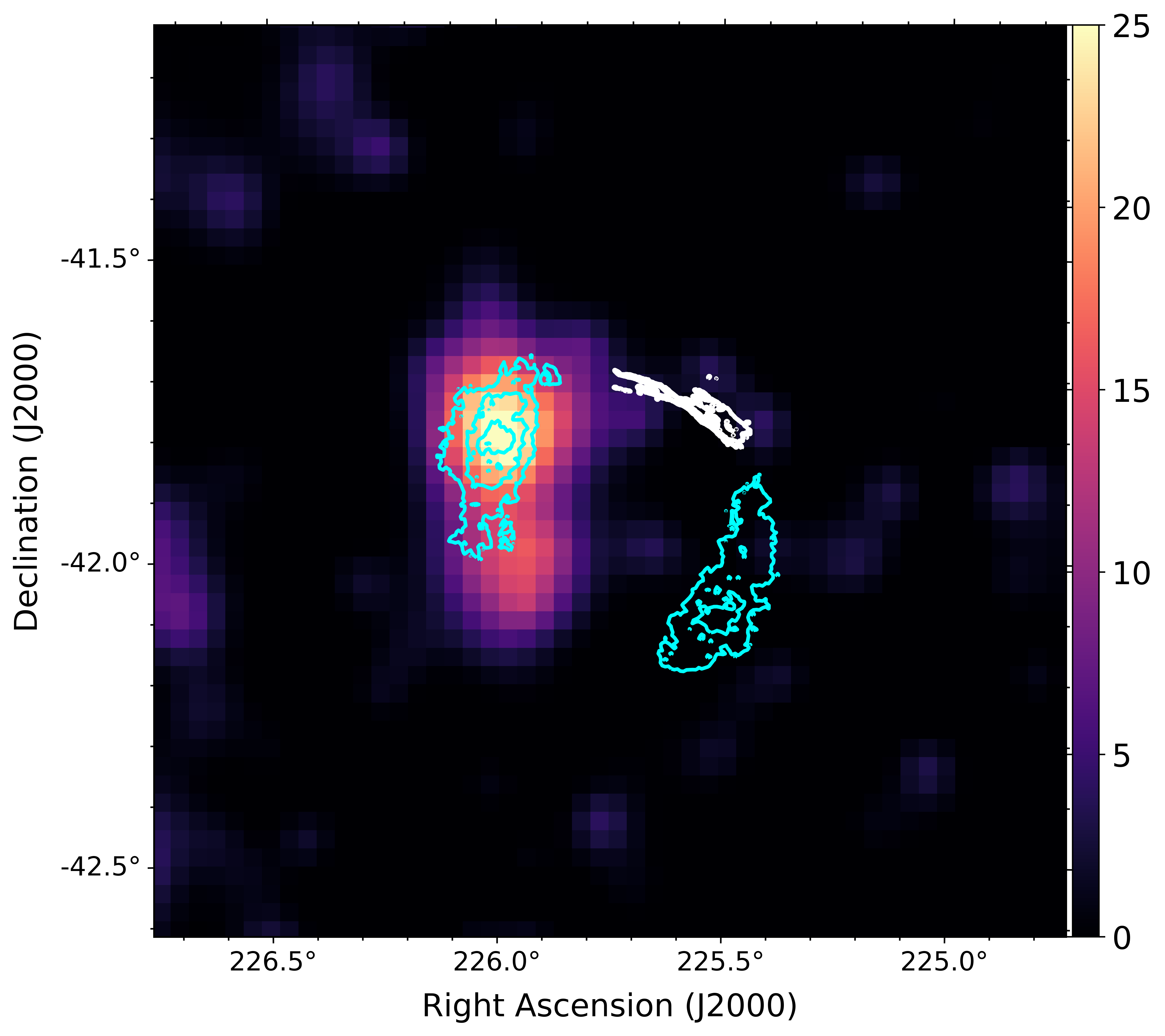}
\includegraphics[width=\linewidth]{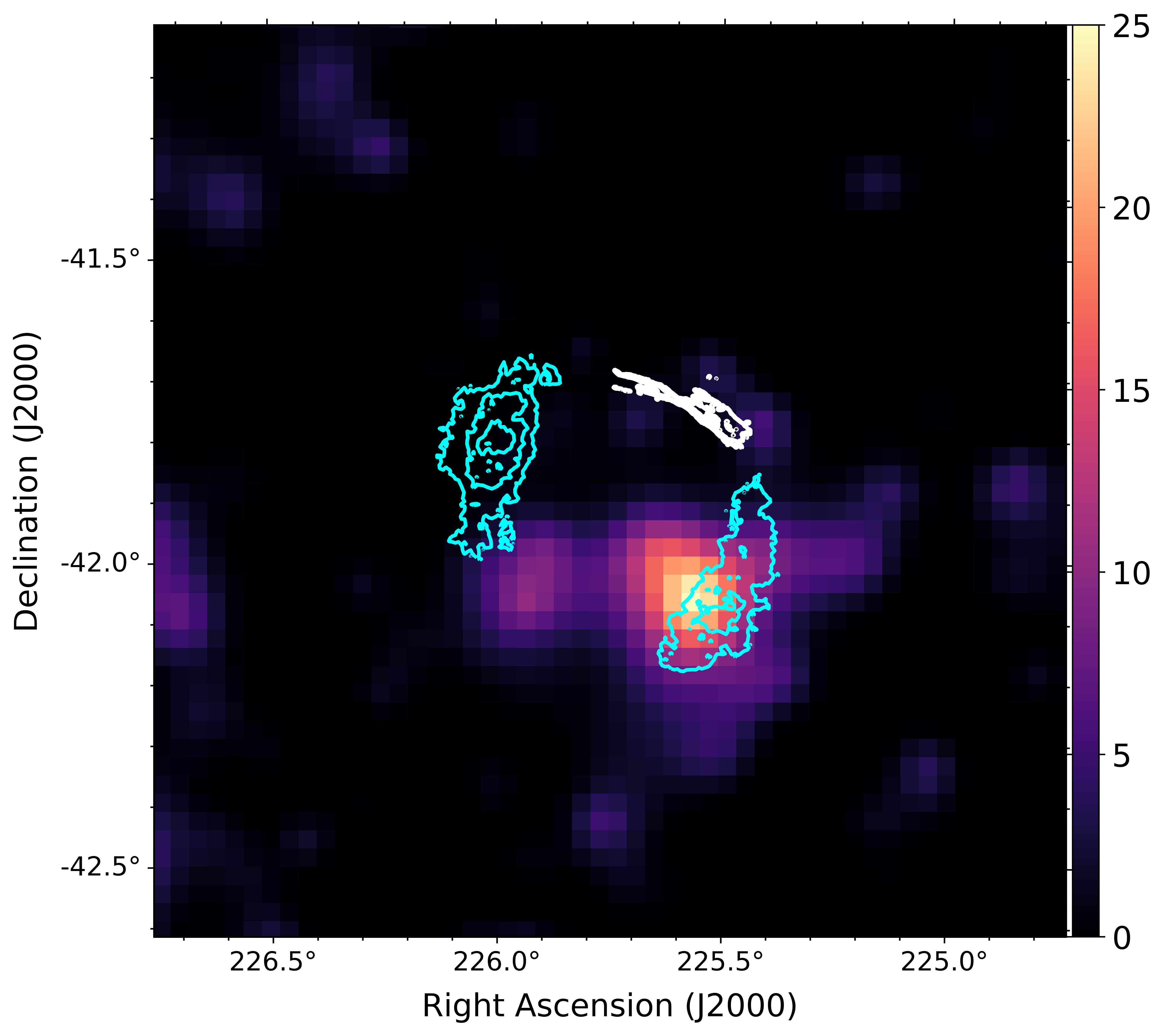}
\includegraphics[width=\linewidth]{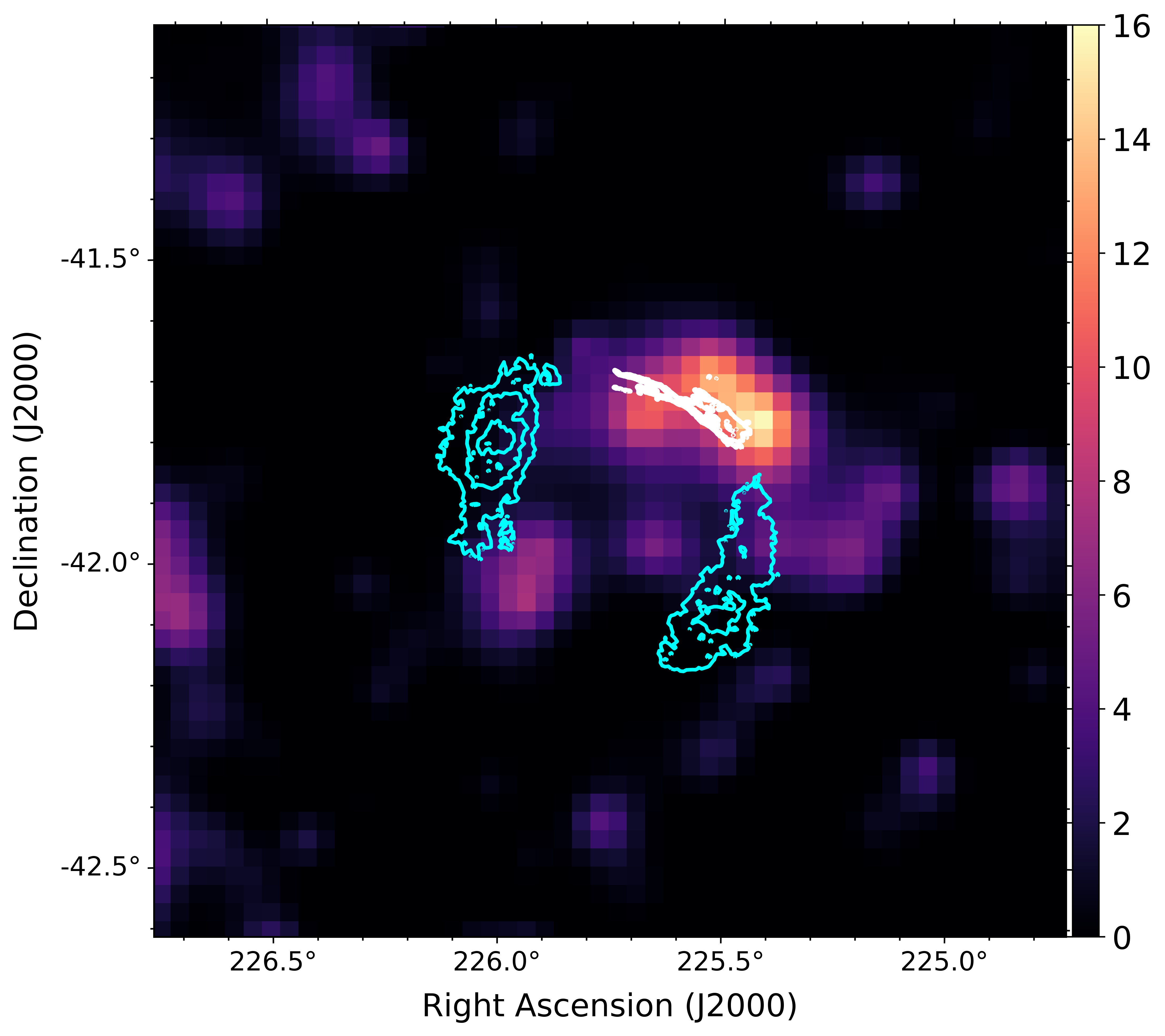}
\caption{\emph{Fermi}-LAT TS map above 1 GeV after removing one of the three components from the best spatial model (Model 6, see Table~\ref{tab:aic}): the H.E.S.S. NE limb (Top panel), the Point source in the SW (Middle panel), the H$\alpha$ component (Bottom panel) with reduced color scale. For all TS maps, we present the $1.5 \degr \times 1.5 \degr$ region of interest around SN 1006. The cyan contours represent the H.E.S.S. significance contours at 3, 5, 7$\sigma$ \citep{2010A&A...516A..62A}. White contours show the H$\alpha$ template used in our analysis.}
\label{fig:tsmaps}
\end{figure}

\begin{table*}[t!]
\caption{Results of the fit of the LAT data between 1 GeV and 1 TeV using different spatial models. The second column reports the likelihood values obtained for each spatial model, while column 3 indicates the number of degrees of freedom adjusted in the model. The delta Akaike criterion, defined as $\Delta$AIC = AIC$_{1}$ - AIC$_i$ = $2 \times (\Delta$ d.o.f. - $\Delta \rm{\ln \mathcal{L}})$, is reported in the fourth column. See Section~\ref{sec:morpho} for more details.}
\label{tab:aic} 
\centering    
\begin{tabular}{lccc}
\hline \hline
Spatial model (number) & Likelihood & d.o.f. & $\Delta$AIC \\
\hline
(1) 2 Point sources & $-$678288.4 & 8 & 0 \\
(2) Disk     &  $-$678289.2  & 5 & 4.4 \\
(3) H.E.S.S. &  $-$678291.6 & 2 & 5.6 \\
(4) H.E.S.S. (NE) + H.E.S.S. (SW) & $-$678288.5   & 4 & 7.8 \\
(5) H.E.S.S. (NE) + H.E.S.S. (SW) + H$\alpha$ & $-$678283.1   & 6 &  14.6 \\
(6) H.E.S.S. (NE) + PS + H$\alpha$ & $-$678280.2   & 8 & 16.4 \\
(7) Radio      &  $-$678289.7   & 2 & 9.4 \\
(8) Radio (2 halves) &  $-$678285.9 & 4 &  13.0 \\
(9) Radio + H.E.S.S. (NE) + H.E.S.S. (SW) & $-$678282.3   & 6 & 16.2 \\
(10) H.E.S.S. (NE) + AGN + H$\alpha$ & $-$678283.9   & 6 & 13.0 \\
(11) H.E.S.S. (NE) + (hadr.+IC)  + H$\alpha$ & $-$678284.2   & 6 & 12.4 \\
(12) \emph{Chandra} (NE) + \emph{Chandra} (SW)  + H$\alpha$ & $-$678284.4   & 6 & 12.0 \\
\end{tabular}
\end{table*}

\section{Morphological analysis of the LAT data} \label{sec:morpho}
The morphological analysis is done between 1 GeV and 1 TeV to take advantage of the improved point spread function (PSF)\footnote{\url{http://www.slac.stanford.edu/exp/glast/groups/canda/lat_Performance.htm}} (with a 68\% containment radius smaller than $0.2^{\circ}$ above 10 GeV). More details on the LAT PSF associated with our configuration are provided in Appendix~\ref{sec:ap} in which we simulate a point source with a spectral index of 2.0. This simulation is added as an inset of Figure~\ref{fig:tsmap1} to better visualize structures larger than our PSF.\\
We perform a binned analysis using all event types\footnote{\url{https://fermi.gsfc.nasa.gov/ssc/data/analysis/documentation/Cicerone/Cicerone_Data/LAT_DP.html}} (PSF0, PSF1, PSF2 and PSF3) with spatial bins of $0.03^{\circ}$. As a first step, the spectral parameters of the sources located up to $3^{\circ}$ from the center of the region of interest (ROI) are fit simultaneously with the Galactic and isotropic diffuse emissions. During this procedure, the 4FGL-DR4 power-law spectral model of the point source 4FGL J1503.6$-$4146 is used to reproduce the $\gamma$-ray emission of the SNR SN 1006. This source is coincident with the NE limb of the SNR and the only one included in the 4FGL-DR4 catalog located within a $1^{\circ}$ radius from the center of the SNR. To search for additional sources in the ROI, we compute a test statistic (TS) map that tests at each pixel the significance of a source with a generic E$^{-2}$ spectrum against the null hypothesis: ${\rm TS}=2(\ln \mathcal{L}_1 - \ln \mathcal{L}_0)$, where $\mathcal{L}_0$ and $\mathcal{L}_1$ are the likelihoods of the background (null hypothesis) and the hypothesis being tested (source plus background). We iteratively add two point sources in the model where the TS exceeded 25. We localize the two additional sources (RA$_{\rm J2000}$, Dec$_{\rm J2000}$ = $223.36^{\circ} \pm 0.03^{\circ}, -45.62^{\circ} \pm 0.02^{\circ}$; $225.58^{\circ} \pm 0.03^{\circ}, -42.04^{\circ} \pm 0.03^{\circ}$) and we fit their power-law spectral parameters. The second point source is coincident with the SW limb as can be seen in Figure~\ref{fig:tsmap1} and Figure~\ref{fig:tsmaps} (middle), proving that the southern part of the SNR is now detected by the LAT with a TS value exceeding 25 assuming a point source hypothesis. We also calculated the improvement assuming extended Gaussians for these two additional sources and obtained $TS_{\rm ext} = 0$ for the first source and 12 (below the threshold of 16 to claim for an extension) for the source coincident with the SW limb which is called PS in the following.
We then perform the morphological analysis of the SNR SN 1006. In each step, we replace the point sources associated with the NE and SW limb of the emission with different geometrical and multi-wavelength templates, fitting both the morphological and spectral parameters of the new components. The results of all morphological tests performed are reported in Table~\ref{tab:aic}. Since we cannot use the likelihood ratio test to compare models that are not nested, we use the Akaike Information Criterion \cite[AIC]{Akaike1998}. We calculate $\Delta$AIC = AIC$_{\mathrm{2 \, point \, sources}}$ - AIC$_i$ = $2 \times (\Delta$ d.o.f. - $\Delta \rm{\ln \mathcal{L}})$ to compare the different models. The different steps of the procedure are the following: fitting a disk, replacing the disk by the H.E.S.S. spatial template \citep{2010A&A...516A..62A}, or using a H.E.S.S. template for each limb separately. This last step provides an excellent fit to the data though some residuals are still apparent on the NW of the SNR, coincident with the bright H$\alpha$ filament, as can be seen in Figure~\ref{fig:tsmaps} (right). The significance of this emission is tested by using an H$\alpha$ spatial template generated from 4m Blanco telescope observations at CTIO \citep{2014ApJ...781...65W} in addition to the NE and SW H.E.S.S. templates, providing a better $\Delta$AIC value (Model 5). Despite the larger number of degrees of freedom (d.o.f.)), the best $\Delta$AIC value is provided with a spatial model replacing the SW H.E.S.S. spatial template by the point source PS J1502.2$-$4203 (Model 6). This tends to demonstrate that the emission detected by the LAT does not correlate perfectly with the one observed at tera-electronvolt energies, despite a $3\sigma$ indication for extension, and might have another origin. This will be discussed further in Section~\ref{sec:model}.\\
We also test a radio spatial template (Model 7) using observations from the GLEAM survey performed with the Murchison Widefield Array (MWA) between 170 and 230 MHz \citep{2017MNRAS.464.1146H}. This spatial template provides on its own an excellent fit to the data as can be seen from the log-likelihood value in in Table~\ref{tab:aic}. Then, dividing the radio spatial template to fit independently the NE and SW limb further improves the quality of the fit. In this case, the associated spectral indices differ by $2.7\sigma$ (the NE limb being harder than the SW), confirming the previous indication of asymmetry derived by \cite{Condon_2017}. However, the best spatial model remains Model~6 even when fitting the NE and SW H.E.S.S. components in addition to the radio template.\\ 
Finally, focusing on the southern point source, we fix its position at the coordinates of the nearby active galactic nucleus (AGN) candidate detected using \emph{NuSTAR} observations (SW point source 2 located at RA$_{\rm J2000}$ = $225.51^{\circ}$, Dec$_{\rm J2000}$ = $-42.03^{\circ}$) by \cite{Li_2018}, which degrades the likelihood by 3.7 with respect to Model~6 in which the position of the point source is free. This spatial model (Model~10) is therefore not favoured. We also test the synthetic (hadronic + leptonic) monochromatic emission of the southwestern limb of SN 1006 at 3 GeV derived for a spherically symmetric interstellar cloud by \cite{Miceli:2016sml} instead of the point source. This Model~11 degrades the likelihood of the fit, demonstrating that the $\gamma$-ray emission might be more complex than foreseen. A similar degradation is seen when using \emph{Chandra} X-ray templates between 2.5 and 7 keV instead of the H.E.S.S. ones for the NE and SW limbs (Model 12). \\   
Table~\ref{tab:morpho} summarizes the morphological parameters of the best fit spatial model (Model~6) showing that the H$\alpha$ component is detected at only 3.3 $\sigma$ assuming two degrees of freedom. The radio template divided in two halves (Model 8) is the spatial model that best fits the data with only two components or fewer (4 degrees of freedom only instead of 6). In this case, the two halves are significantly detected with a TS of 40 for the NE side (6$\sigma$ for 2 d.o.f.) and 34 for the SW (5.5$\sigma$ for 2 d.o.f.).

\begin{table*}
\caption{\emph{Fermi}-LAT morphological parameters of the three sources in Model 6 and two sources in Model 8 derived between 1 GeV and 1 TeV.}
\label{tab:morpho}
\centering    
\begin{tabular}{llcccc}
\hline \hline
Model name & Source name & RA & Dec & TS \\
 & & ($^{\circ}$) & ($^{\circ}$) &  \\ 
\hline \hline
Model 6 & H.E.S.S. (NE)      &    &   & 36 \\
 & PS J1502.2$-$4203 (PS) & 225.57 & $-$42.06 & 26 \\
& H$\alpha$     &  &   & 14 \\
\hline
Model 8 & Radio (NE) & & & 40 \\
 & Radio (SW) & & & 34 \\
\end{tabular}
\end{table*}

\begin{table*}
\caption{\emph{Fermi}-LAT spectral parameters of the components in Model 6 and Model 8 between 100 MeV and 1 TeV. The first (second) error represent statistical (systematic) error respectively. Columns 4 and 5 provide the TS value and the improvement of the log-normal representation with respect to the PL model TS$_{LP}$.}
\label{tab:spec}
\centering    
\begin{tabular}{llccccc}
\hline \hline
Model name & Source name & Spectral index & Energy flux (100 MeV - 1 TeV) & TS & TS$_{LP}$ \\
 & & & ($10^{-12}$ erg cm$^{-2}$ s$^{-1}$) & \\ 
\hline
Model 6 & H.E.S.S. (NE)        & $1.7 \pm 0.1 \pm 0.1$ & $2.9 \pm 0.3 \pm 0.2$ &  35 & 0.0 \\
 & PS J1502.2$-$4203       & $2.1 \pm 0.1 \pm 0.1$ & 1.6 $\pm 0.3 \pm 0.2$ &  26 & 1.5 \\
 & H$\alpha$      & $ 2.1 \pm 0.2 \pm 0.3$ & $1.9 \pm 0.3 \pm 0.6$ & 14  & 0.2 \\
\hline
Model 8 & Radio (NE)        & $1.7 \pm 0.1 \pm 0.1$ & $3.6 \pm 0.2 \pm 0.2$ &  43 & 0.2 \\
 & Radio (SW)       & $2.2 \pm 0.1 \pm 0.1$ & 2.2 $\pm 0.2 \pm 0.2$ &  33 & 2.2 \\ 
\end{tabular}
\end{table*}

\section{Spectral analysis of the LAT data} \label{sec:spec}
\begin{figure}[ht]
\includegraphics[width=0.99\linewidth]{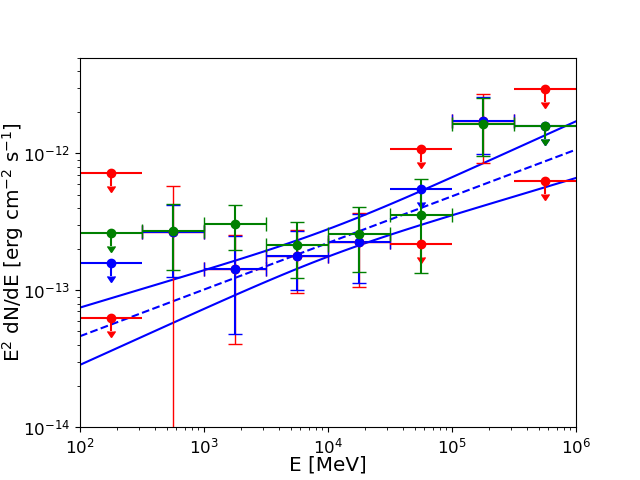}
\includegraphics[width=0.99\linewidth]{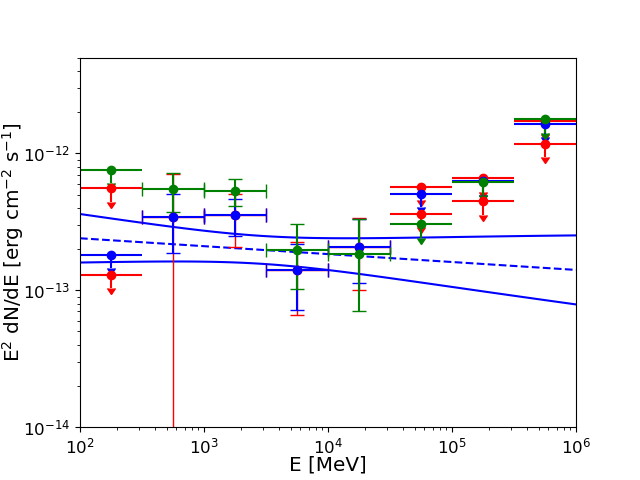}
\hspace{0.51\textwidth}
\includegraphics[width=0.99\linewidth]{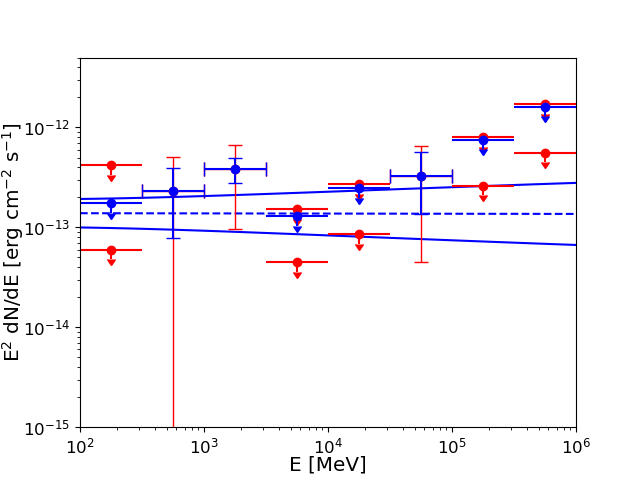}
\caption{\emph{Fermi}-LAT spectral energy distribution of the three components included in the best spatial model (Model 6): the H.E.S.S. NE limb (Top panel), the Point source coincident with the SW limb (Middle panel), the H$\alpha$ component (Bottom panel). For all SEDs, blue error bars represent the statistical uncertainties, while the red ones correspond to the statistical and systematic uncertainties added in quadrature. For upper limits, the two red arrows indicate the extrema of upper limits obtained with the different systematics. The solid and dashed blue lines represent the best spectral fit and its 68\% confidence band. For the NE and SW components, the green spectral points indicate the spectra derived for Model~8 using the two radio halves (only statistical errors are presented for better visibility; systematic uncertainties for this model are indicated in Table~\ref{tab:spec}).}
\label{fig:seds}
\end{figure}

 We perform the spectral analysis from 100 MeV to 1 TeV with a summed likelihood method to simultaneously fit events with different angular reconstruction quality (PSF0 to PSF3 event types). To ensure that our results are not affected by the spatial model assumed, the spectral analysis is performed assuming the best-fit spatial template with 3 components and with 2 components (Model 6 and Model 8, see Table~\ref{tab:morpho}). This summed likelihood method was used in several \emph{Fermi}-LAT analyses including the Kepler SNR \citep{2022A&A...660A.129A} allowing a more sensitive analysis.
 We use PSF1, PSF2 and PSF3 events below 1 GeV, and all event types above 1 GeV. Since one additional year of data is used with respect to the 4FGL-DR4 catalog, we first check whether additional sources are needed in the model by examining the TS maps above 100 MeV. Three additional sources are detected at the following positions RA$_{\rm J2000}$, Dec$_{\rm J2000}$ = ($226.88^{\circ}, -43.20^{\circ}$); ($231.30^{\circ}, -43.77^{\circ}$); ($218.88^{\circ}, -39.34^{\circ}$) with TS values of 32, 27 and 25, respectively. These sources are not detected significantly in the 1 GeV -- 1 TeV range used in Section~\ref{sec:morpho} and are all located over $1.5^{\circ}$ from the center of our region of interest.\\
Adding these three point sources in our model, we then test a simple power-law model and a logarithmic parabola for the three (or two, depending on the spatial model assumed) components reproducing the $\gamma$-ray emission of SN~1006. During this procedure, the spectral parameters of sources located up to $3^{\circ}$ from the ROI center are again left free during the fit, like those of the Galactic and isotropic diffuse emissions. The improvement between the power-law model and the logarithmic parabola is tested using the likelihood ratio test (TS$_{LP}$ in Table~\ref{tab:spec}) and is not significant for any of the component with the current statistics.\\ 
We finally derive the \emph{Fermi}-LAT spectral points (spectral energy distribution or SED) for the three components of the SNR SN 1006, shown in Figure~\ref{fig:seds}, by dividing the 100 MeV -- 1 TeV energy range into 8 logarithmically-spaced energy bins and  performing a maximum likelihood spectral analysis to estimate the photon flux in each interval, assuming a power-law shape with fixed photon index $\Gamma$=2 for the source of interest. The normalizations of the diffuse Galactic and isotropic emission are left free in each energy bin as well as those of the sources within $3^{\circ}$. A 95\% confidence level upper limit is computed when the TS value is lower than $4$. Figure~\ref{fig:seds} also presents the spectral points derived for the two components of Model~8 (the NE and SW radio halves, in green) which are in very good agreement with those derived with the three component model within statistical errors.\\
We then estimate the systematic errors on the spectral parameters of the three and the two-component models. They depend on the uncertainties on the Galactic diffuse emission model, on the effective area, and on the spatial shape of the source. The first is calculated using eight alternative diffuse emission models following the same procedure as in the first \emph{Fermi}-LAT supernova remnant catalog \citep{2016ApJS..224....8A} and the second is obtained by applying two scaling functions on the effective area following the standard method defined in \cite{2012ApJS..203....4A}. Finally, we consider the impact on the spectral parameters of the NE and SW component when changing the spatial model. These different sources of systematic uncertainties are added in quadrature to represent the total systematic uncertainty. The spectral parameters of the three components of Model 6 and the two components of Model 8 are presented in Table~\ref{tab:spec}, together with their estimated systematic errors.

\section{Discussion and modeling of the multi-wavelength data}
\label{sec:model}
Located 550 pc above the plane (assuming a distance of 2.2 kpc), the SNR SN 1006 evolves in a tenuous environment and its shock velocity exceeds 5000 km~s$^{-1}$ \citep[]{Katsuda2009, 2014ApJ...781...65W}. Deep X-ray observations through a dedicated \emph{XMM-Newton} Large Programme have revealed that the ambient density is $\sim 0.035$ cm$^{-3}$ in the south-eastern (SE) limb \citep{2012A&A...546A..66M}. Similar estimates have been obtained within the \emph{Chandra} Large Programme \citep{2014ApJ...781...65W, 2022NatCo..13.5098G} showing an homogeneous spatial distribution of the ambient medium around SN 1006 in the SE and up to the NE regions. \cite{2022ApJ...933..157S} recently proposed a different description of the circumstellar medium but there exist no indications of shocked gas at those high densities. The tenuous environment surrounding the remnant does not favor the proton–proton interactions to produce the $\gamma$-ray emission detected by the LAT, especially in the NE limb. However, \cite{2014ApJ...782L..33M} have revealed a dense atomic cloud interacting with the SW synchrotron rim of SN 1006, where efficient particle acceleration is at work. Our assumption is therefore that the observed giga-electronvolt $\gamma$-ray emission in the NE is dominated by the inverse Compton emission of accelerated electrons, explaining the excellent correlation between the X-ray synchrotron emission, the tera-electronvolt emission detected by H.E.S.S. and the signal detected by the LAT. On the other hand, the emission observed by the LAT in the SW would be mostly of hadronic nature (proton-proton interactions). This might also be the case in the NW region of the remnant where a bright H$\alpha$ filament is detected due to the slowing down of the shock by interaction with dense material. However, the emission detected by the LAT in this region is too faint to provide strong constraints. We will thus concentrate on the modeling of the two bright NE and SW rims using the spectral points derived in Model~8 (two radio halves), because they are directly comparable to the radio data, and the electrons emitting inverse Compton in the GeV range and synchrotron in the GHz range have similar energies. \\
For simplicity, we try to model each limb separately using identical parameters except the density of the medium to determine whether the increase in density in the SW could explain the different emission detected by \emph{Fermi}. To this aim, we use the radio \citep{2001ApJ...558..739A} and X-ray \citep{2008PASJ...60S.153B} data from the whole remnant. At radio energies, we estimate the fraction of the flux in each component to be $\approx 50$\% of the whole SNR flux using observations from the GLEAM survey. Doing the same for the X-ray synchrotron flux between 2.5 and 7 keV, we derive a fraction of 57\% of the whole SNR flux in the NE component and 43\% in the SW part. We use these fractions in our modeling of each component. We also take into account the tera-electronvolt emission detected in the NE and SW limbs by H.E.S.S. \citep{2010A&A...516A..62A}.\\ 
In each limb, we adopt the simplest possible assumption that all emission originates from a single population of accelerated protons and electrons contained in a region characterized by a constant matter density and magnetic field strength. The particle spectra are assumed to follow a power-law with an exponential cutoff $\mathrm dN/\mathrm dE \propto \eta_{e, p} \, E^{-\Gamma} \exp(-E/E_{\rm max})$, with the same injection index for both electrons and protons set at 2.2 to reproduce the radio spectral index of $\approx 0.6$. The cut-off energies for electrons and protons are different, allowing a lower energy cut-off for electrons due to synchrotron losses. The radiative models from the \textit{naima} packages \citep{2015ICRC...34..922Z} have been used with the \textit{Pythia8} parametrization of  \citet{2014PhRvD..90l3014K} for the $\pi^{o}$ decay. In each limb, we assume an electron-to-proton ratio $K_{ep} = \eta_e/\eta_p = 0.01$ and a distance of 2.2 kpc. One should note that our model assumes simple diffuse shock acceleration though \cite{Cristofari:2019pqh} have demonstrated that re-acceleration of particles at SNR shocks can play a significant role, especially at low gas density. That would change little our modeling because re-acceleration predicts the same spectral slope for electrons and protons.

\begin{figure*}[t!]

\includegraphics[width=0.5\textwidth]{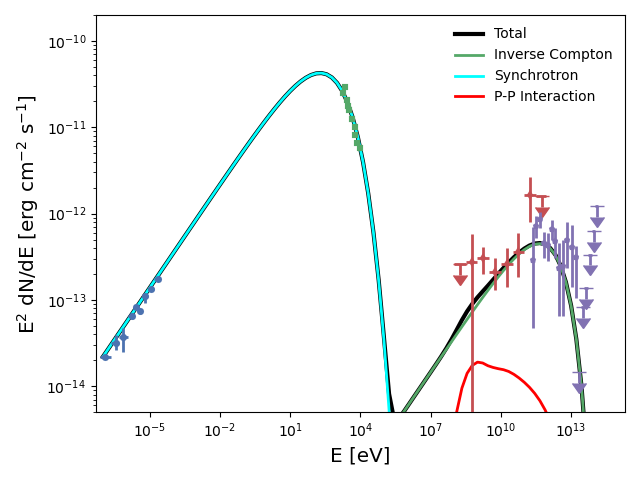}
\includegraphics[width=0.5\textwidth]{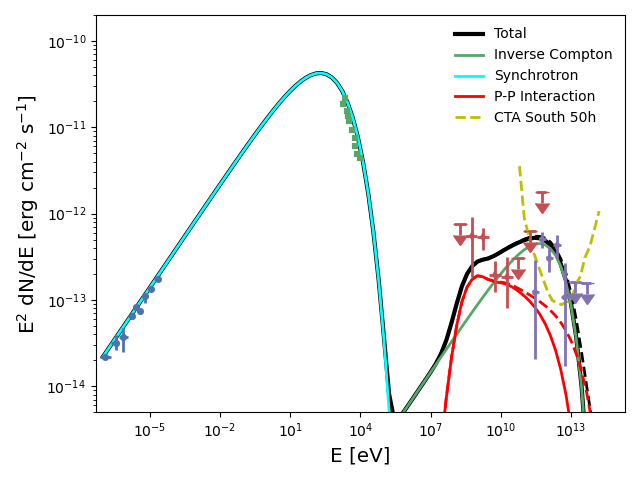}

\caption{Spectral energy modeling of the NE (Left) and SW (Right) regions of SN 1006. For all experiments except \emph{Fermi}, only statistical errors are shown. Modeling parameters are listed in Table \ref{tab:naima}. The cyan and green lines represent the synchrotron emission and the IC emission, respectively. The black and red dotted lines represent the total and pion decay emission derived for a proton energy cut-off at 200 TeV while the solid ones are derived for 20 TeV. The dashed yellow line indicates the sensitivity of CTA for 50 hours of observation (latest response function: Southern array Prod 5). The blue radio \citep{2001ApJ...558..739A} and green X-ray \citep{2008PASJ...60S.153B} data from the whole SNR have been scaled for each limb (see Section~\ref{sec:model} for more details). The H.E.S.S. spectral points for each limb \citep{2010A&A...516A..62A} are indicated in purple. The data points derived in this analysis for the NE and SW limbs are presented in red.}
\label{fig:naima}
\end{figure*}

Modeled SEDs are presented in Figure~\ref{fig:naima} and their associated parameters are given in Table~\ref{tab:naima}. While the chosen parameters are not unique in their ability to fit the broadband spectrum, they are compatible with those considered in previous works \citep[]{2015A&A...580A..74A, Xing2016} and they demonstrate that the difference in spectrum and morphology detected in the SW limb by the LAT could well be explained by a simple difference in density from an average value of 0.035 cm$^{-3}$ in the NE to 0.35 cm$^{-3}$ in the SW. In this case, the total energy transferred to accelerated protons is on the order of $2.6 \times 10^{49}$~erg. This energy is concentrated in the limbs and would amount to 13\% of the kinetic energy, assuming a volume factor of $20$\% and an explosion energy of $10^{51}$ erg. This is consistent with previous estimates in the limbs by \cite{2022NatCo..13.5098G} and with the value derived for the historical supernova Kepler ~\citep{2022A&A...660A.129A}. Our new LAT analysis and simple modeling thus confirms the magnetohydrodynamic simulations performed by \cite{Miceli:2016sml}. However, if the SW limb is dominated by gamma rays generated by proton-proton interaction, one would expect a correlation with the density of the medium and more precisely with the atomic cloud detected by \cite{2014ApJ...782L..33M}. This does not seem to be the case here, though more statistics would be needed to characterize the morphology of the southwestern emission detected by the LAT. Similarly, we note some hint for a possible double-peaked feature in the SW spectrum, which is not statistically significant. More statistics with the LAT but also at tera-electronvolt energies with the future Cherenkov Telescope Array Observatory (CTAO) would be needed to confirm this effect. One can note that the modeling proposed here is in good agreement with the mixed scenario discussed in \cite{2019PASJ...71...77X}. They found that the contribution of the leptonic component is only <30\% lower than that of their purely leptonic model and that $~\sim 2$\% of the kinetic energy of the SNR was converted into hadrons to fit the multi-wavelength data of the SW limb. In our case as well, the gamma-ray emission is dominated by the leptonic component above a few GeV and the hadronic component only dominates at low energy.\\ 
Since the cut-off energy for protons cannot be constrained with the current multi-wavelength data, two different values (20 TeV and 200 TeV) are tested for the SW limb in which $\gamma$-ray emission produced by proton-proton interaction is enhanced due to the higher density. As can be seen in Figure~\ref{fig:naima}, an observation time larger than 50 hours will be needed for CTAO\footnote{The sensitivities for CTAO are available at \url{https://www.ctao.org/for-scientists/performance/}} to be able to constrain the high energy cut-off of the accelerated protons. In both limbs, the magnetic field value is set at 30~$\mu$G, constrained by the X-ray to TeV flux ratio. This value is well below the magnetic field value required to explain the very thin X-ray filaments in the NE rim of SN 1006 \citep{2003ApJ...586.1162L}. In addition, while our simple one-zone model can account for the measured $\gamma $-ray flux, it fails to reproduce the highest energy spectral points detected by H.E.S.S. in the NE limb, which are higher than the expectations from the inverse Compton (IC) process as already noted by \citet{2010A&A...516A..62A}. Such precise modeling at the highest energies would require deriving the X-ray spectrum on the exact same region analyzed with \emph{Fermi}-LAT. Thanks to the careful scaling performed here on the X-ray spectrum of the whole SNR, we do not expect significant differences on the main parameters of the modeling except the maximum energy of the accelerated electrons.\\ 

\section{Conclusion} \label{sec:conclu}
By using 15 years of \emph{Fermi}-LAT data and a summed likelihood analysis with the PSF event types, we are able to perform a complete morphological study of the gamma-ray emission of the SNR SN 1006. We significantly detect both the NE and SW rims of the SNR with TS values of 40 (6$\sigma$ for 2 degrees of freedom) and 34 (5.5$\sigma$ for 2 degrees of freedom) respectively, using radio templates using GLEAM survey data from MWA observations. Additionally, our analysis reveals a $3\sigma$ excess coincident with the bright NW H$\alpha$ filament. We can confirm the harder spectrum of the NE rim with respect to the SW rim. This asymmetry can be well reproduced with a simple one-zone model in which the only difference is the gas density an order of magnitude higher in the SW than in the NE, thus enhancing the gamma-ray emission produced by proton-proton interaction in the SW especially below a few GeV. In contrast, the gamma-ray emission in the NE is dominated by accelerated electrons radiating through inverse Compton processes. Assuming a gas density of 0.035 cm$^{-3}$ in the NE to 0.35 cm$^{-3}$ in the SW, the total energy transferred to accelerated protons is on the order of $2.6 \times 10^{49}$~erg. While the cut-off energy is well constrained, especially by X-ray data, an observation time larger than 50 hours would be needed for the future Cherenkov Telescope Array Observatory to be able to constrain the high energy cut-off of the accelerated protons.\\

\textit{Acknowledgments}:
The \textit{Fermi} LAT Collaboration acknowledges generous ongoing support
from a number of agencies and institutes that have supported both the
development and the operation of the LAT as well as scientific data analysis.
These include the National Aeronautics and Space Administration and the
Department of Energy in the United States, the Commissariat \`a l'Energie Atomique
and the Centre National de la Recherche Scientifique / Institut National de Physique
Nucl\'eaire et de Physique des Particules in France, the Agenzia Spaziale Italiana
and the Istituto Nazionale di Fisica Nucleare in Italy, the Ministry of Education,
Culture, Sports, Science and Technology (MEXT), High Energy Accelerator Research
Organization (KEK) and Japan Aerospace Exploration Agency (JAXA) in Japan, and
the K.~A.~Wallenberg Foundation, the Swedish Research Council and the
Swedish National Space Board in Sweden.
 Additional support for science analysis during the operations phase is gratefully
acknowledged from the Istituto Nazionale di Astrofisica in Italy and the Centre
National d'\'Etudes Spatiales in France. This work performed in part under DOE
Contract DE-AC02-76SF00515.\\
MLG acknowledges support from the Alexander von Humboldt Foundation and from ANR for the GAMALO project under reference ANR-19-CE31-0014.

\textit{Softwares}: 
This research made use of the Python package \textit{fermipy} for the \emph{Fermi}-LAT analysis \citep{2017ICRC...35..824W}. The \textit{naima} package was used for the modeling \citep{2015ICRC...34..922Z}.

\begin{table*}

\caption{List of parameters obtained from the modeling of the spectral energy distribution in the NE and SW limbs: magnetic field (column 2), ambient density (column 3), injection index (column 4), electron maximum energy (column 5), proton maximum energy (column 6), energy injected in protons (column 7) and electron-to-proton ratio (column 8). Parameters in brackets are fixed from observables, while other parameters are adjusted to the data. The energy budget values are integrated above 1 GeV. Two cut-off energies for protons are tested as can be seen in Figure~\ref{fig:naima}.}

\begin{tabular}{lcccccccc}
\hline \hline
Limb & $B$ & $n_0$ & $\Gamma$ & $E_{\rm max,e}$ & $E_{\rm max,p}$ & $W_{p}$ & $K_{ep}$ \\
& $\mu$G & cm$^{-3}$ &  & TeV & TeV & erg & \\
 \hline
NE & 30 & 0.035 & [2.2] & 15 & 20 / (200) & 1.3$\times 10^{49}$  & 0.01\\
SW  & 30 & 0.35 & [2.2] & 15 & 20 / (200) & 1.3$\times 10^{49}$ & 0.01 \\
\end{tabular}

\label{tab:naima}
\end{table*}

\bibliography{biblio}{}

\begin{thebibliography}{30}
\expandafter\ifx\csname natexlab\endcsname\relax\def\natexlab#1{#1}\fi

\bibitem[{{Acero} {et~al.}(2016){Acero}, {Ackermann}, {Ajello}, {Baldini}, {Ballet}, {Barbiellini}, {Bastieri}, {Bellazzini}, {Bissaldi}, {Blandford}, {Bloom}, {Bonino}, {Bottacini}, {Brandt}, {Bregeon}, {Bruel}, {Buehler}, {Buson}, {Caliandro}, {Cameron}, {Caputo}, {Caragiulo}, {Caraveo}, {Casandjian}, {Cavazzuti}, {Cecchi}, {Chekhtman}, {Chiang}, {Chiaro}, {Ciprini}, {Claus}, {Cohen}, {Cohen-Tanugi}, {Cominsky}, {Condon}, {Conrad}, {Cutini}, {D'Ammando}, {de Angelis}, {de Palma}, {Desiante}, {Digel}, {Di Venere}, {Drell}, {Drlica-Wagner}, {Favuzzi}, {Ferrara}, {Franckowiak}, {Fukazawa}, {Funk}, {Fusco}, {Gargano}, {Gasparrini}, {Giglietto}, {Giommi}, {Giordano}, {Giroletti}, {Glanzman}, {Godfrey}, {Gomez-Vargas}, {Grenier}, {Grondin}, {Guillemot}, {Guiriec}, {Gustafsson}, {Hadasch}, {Harding}, {Hayashida}, {Hays}, {Hewitt}, {Hill}, {Horan}, {Hou}, {Iafrate}, {Jogler}, {J{\'o}hannesson}, {Johnson}, {Kamae}, {Katagiri}, {Kataoka}, {Katsuta}, {Kerr}, {Kn{\"o}dlseder}, {Kocevski}, {Kuss}, {Laffon}, {Lande},
  {Larsson}, {Latronico}, {Lemoine-Goumard}, {Li}, {Li}, {Longo}, {Loparco}, {Lovellette}, {Lubrano}, {Magill}, {Maldera}, {Marelli}, {Mayer}, {Mazziotta}, {Michelson}, {Mitthumsiri}, {Mizuno}, {Moiseev}, {Monzani}, {Moretti}, {Morselli}, {Moskalenko}, {Murgia}, {Nemmen}, {Nuss}, {Ohsugi}, {Omodei}, {Orienti}, {Orlando}, {Ormes}, {Paneque}, {Perkins}, {Pesce-Rollins}, {Petrosian}, {Piron}, {Pivato}, {Porter}, {Rain{\`o}}, {Rando}, {Razzano}, {Razzaque}, {Reimer}, {Reimer}, {Renaud}, {Reposeur}, {Rousseau}, {Saz Parkinson}, {Schmid}, {Schulz}, {Sgr{\`o}}, {Siskind}, {Spada}, {Spandre}, {Spinelli}, {Strong}, {Suson}, {Tajima}, {Takahashi}, {Tanaka}, {Thayer}, {Thompson}, {Tibaldo}, {Tibolla}, {Torres}, {Tosti}, {Troja}, {Uchiyama}, {Vianello}, {Wells}, {Wood}, {Wood}, {Yassine}, {den Hartog}, \& {Zimmer}}]{2016ApJS..224....8A}
{Acero}, F., {Ackermann}, M., {Ajello}, M., {et~al.} 2016, \apjs, 224, 8

\bibitem[{{Acero} {et~al.}(2010){Acero}, {Aharonian}, {Akhperjanian}, {Anton}, {Barres de Almeida}, {Bazer-Bachi}, {Becherini}, {Behera}, {Beilicke}, {Bernl{\"o}hr}, {Bochow}, {Boisson}, {Bolmont}, {Borrel}, {Brucker}, {Brun}, {Brun}, {B{\"u}hler}, {Bulik}, {B{\"u}sching}, {Boutelier}, {Chadwick}, {Charbonnier}, {Chaves}, {Cheesebrough}, {Conrad}, {Chounet}, {Clapson}, {Coignet}, {Dalton}, {Daniel}, {Davids}, {Degrange}, {Deil}, {Dickinson}, {Djannati-Ata{\"\i}}, {Domainko}, {O'C. Drury}, {Dubois}, {Dubus}, {Dyks}, {Dyrda}, {Egberts}, {Eger}, {Espigat}, {Fallon}, {Farnier}, {Fegan}, {Feinstein}, {Fiasson}, {F{\"o}rster}, {Fontaine}, {F{\"u}{\ss}ling}, {Gabici}, {Gallant}, {G{\'e}rard}, {Gerbig}, {Giebels}, {Glicenstein}, {Gl{\"u}ck}, {Goret}, {G{\"o}ring}, {Hauser}, {Hauser}, {Heinz}, {Heinzelmann}, {Henri}, {Hermann}, {Hinton}, {Hoffmann}, {Hofmann}, {Hofverberg}, {Holleran}, {Hoppe}, {Horns}, {Jacholkowska}, {de Jager}, {Jahn}, {Jung}, {Katarzy{\'n}ski}, {Katz}, {Kaufmann}, {Kerschhaggl}, {Khangulyan},
  {Kh{\'e}lifi}, {Keogh}, {Klochkov}, {Klu{\'z}niak}, {Kneiske}, {Komin}, {Kosack}, {Kossakowski}, {Lamanna}, {Lemoine-Goumard}, {Lenain}, {Lohse}, {Marandon}, {Marcowith}, {Masbou}, {Maurin}, {McComb}, {Medina}, {M{\'e}hault}, {Moderski}, {Moulin}, {Naumann-Godo}, {de Naurois}, {Nedbal}, {Nekrassov}, {Nicholas}, {Niemiec}, {Nolan}, {Ohm}, {Olive}, {de O{\~n}a Wilhelmi}, {Orford}, {Ostrowski}, {Panter}, {Paz Arribas}, {Pedaletti}, {Pelletier}, {Petrucci}, {Pita}, {P{\"u}hlhofer}, {Punch}, {Quirrenbach}, {Raubenheimer}, {Raue}, {Rayner}, {Reimer}, {Renaud}, {de Los Reyes}, {Rieger}, {Ripken}, {Rob}, {Rosier-Lees}, {Rowell}, {Rudak}, {Rulten}, {Ruppel}, {Ryde}, {Sahakian}, {Santangelo}, {Schlickeiser}, {Sch{\"o}ck}, {Sch{\"o}nwald}, {Schwanke}, {Schwarzburg}, {Schwemmer}, {Shalchi}, {Sushch}, {Sikora}, {Skilton}, {Sol}, {Stawarz}, {Steenkamp}, {Stegmann}, {Stinzing}, {Superina}, {Szostek}, {Tam}, {Tavernet}, {Terrier}, {Tibolla}, {Tluczykont}, {van Eldik}, {Vasileiadis}, {Venter}, {Venter}, {Vialle}, {Vincent},
  {Vink}, {Vivier}, {V{\"o}lk}, {Volpe}, {Vorobiov}, {Wagner}, {Ward}, {Zdziarski}, {Zech}, \& {H.~E.~S.~S. Collaboration}}]{2010A&A...516A..62A}
{Acero}, F., {Aharonian}, F., {Akhperjanian}, A.~G., {et~al.} 2010, \aap, 516, A62

\bibitem[{{Acero} {et~al.}(2022){Acero}, {Lemoine-Goumard}, \& {Ballet}}]{2022A&A...660A.129A}
{Acero}, F., {Lemoine-Goumard}, M., \& {Ballet}, J. 2022, \aap, 660, A129

\bibitem[{{Acero} {et~al.}(2015){Acero}, {Lemoine-Goumard}, {Renaud}, {Ballet}, {Hewitt}, {Rousseau}, \& {Tanaka}}]{2015A&A...580A..74A}
{Acero}, F., {Lemoine-Goumard}, M., {Renaud}, M., {et~al.} 2015, \aap, 580, A74

\bibitem[{{Ackermann} {et~al.}(2012){Ackermann}, {Ajello}, {Albert}, {Allafort}, {Atwood}, {Axelsson}, {Baldini}, {Ballet}, {Barbiellini}, {Bastieri}, {Bechtol}, {Bellazzini}, {Bissaldi}, {Blandford}, {Bloom}, {Bogart}, {Bonamente}, {Borgland}, {Bottacini}, {Bouvier}, {Brandt}, {Bregeon}, {Brigida}, {Bruel}, {Buehler}, {Burnett}, {Buson}, {Caliandro}, {Cameron}, {Caraveo}, {Casandjian}, {Cavazzuti}, {Cecchi}, {{\c{C}}elik}, {Charles}, {Chaves}, {Chekhtman}, {Cheung}, {Chiang}, {Ciprini}, {Claus}, {Cohen-Tanugi}, {Conrad}, {Corbet}, {Cutini}, {D'Ammando}, {Davis}, {de Angelis}, {DeKlotz}, {de Palma}, {Dermer}, {Digel}, {Silva}, {Drell}, {Drlica-Wagner}, {Dubois}, {Favuzzi}, {Fegan}, {Ferrara}, {Focke}, {Fortin}, {Fukazawa}, {Funk}, {Fusco}, {Gargano}, {Gasparrini}, {Gehrels}, {Giebels}, {Giglietto}, {Giordano}, {Giroletti}, {Glanzman}, {Godfrey}, {Grenier}, {Grove}, {Guiriec}, {Hadasch}, {Hayashida}, {Hays}, {Horan}, {Hou}, {Hughes}, {Jackson}, {Jogler}, {J{\'o}hannesson}, {Johnson}, {Johnson}, {Johnson},
  {Kamae}, {Katagiri}, {Kataoka}, {Kerr}, {Kn{\"o}dlseder}, {Kuss}, {Lande}, {Larsson}, {Latronico}, {Lavalley}, {Lemoine-Goumard}, {Longo}, {Loparco}, {Lott}, {Lovellette}, {Lubrano}, {Mazziotta}, {McConville}, {McEnery}, {Mehault}, {Michelson}, {Mitthumsiri}, {Mizuno}, {Moiseev}, {Monte}, {Monzani}, {Morselli}, {Moskalenko}, {Murgia}, {Naumann-Godo}, {Nemmen}, {Nishino}, {Norris}, {Nuss}, {Ohno}, {Ohsugi}, {Okumura}, {Omodei}, {Orienti}, {Orlando}, {Ormes}, {Paneque}, {Panetta}, {Perkins}, {Pesce-Rollins}, {Pierbattista}, {Piron}, {Pivato}, {Porter}, {Racusin}, {Rain{\`o}}, {Rando}, {Razzano}, {Razzaque}, {Reimer}, {Reimer}, {Reposeur}, {Reyes}, {Ritz}, {Rochester}, {Romoli}, {Roth}, {Sadrozinski}, {Sanchez}, {Saz Parkinson}, {Sbarra}, {Scargle}, {Sgr{\`o}}, {Siegal-Gaskins}, {Siskind}, {Spandre}, {Spinelli}, {Stephens}, {Suson}, {Tajima}, {Takahashi}, {Tanaka}, {Thayer}, {Thayer}, {Thompson}, {Tibaldo}, {Tinivella}, {Tosti}, {Troja}, {Usher}, {Vandenbroucke}, {Van Klaveren}, {Vasileiou}, {Vianello},
  {Vitale}, {Waite}, {Wallace}, {Winer}, {Wood}, {Wood}, {Wood}, {Yang}, \& {Zimmer}}]{2012ApJS..203....4A}
{Ackermann}, M., {Ajello}, M., {Albert}, A., {et~al.} 2012, \apjs, 203, 4

\bibitem[{Akaike(1998)}]{Akaike1998}
Akaike, H. 1998, Information Theory and an Extension of the Maximum Likelihood Principle, ed. E.~Parzen, K.~Tanabe, \& G.~Kitagawa (New York, NY: Springer New York), 199--213

\bibitem[{{Allen} {et~al.}(2001){Allen}, {Petre}, \& {Gotthelf}}]{2001ApJ...558..739A}
{Allen}, G.~E., {Petre}, R., \& {Gotthelf}, E.~V. 2001, \apj, 558, 739

\bibitem[{Araya \& Frutos(2012)}]{Araya2012}
Araya, M. \& Frutos, F. 2012, Monthly Notices of the Royal Astronomical Society, 425, 2810

\bibitem[{{Atwood} {et~al.}(2009){Atwood}, {Abdo}, {Ackermann}, {Althouse}, {Anderson}, {Axelsson}, {Baldini}, {Ballet}, {Band}, {Barbiellini}, {Bartelt}, {Bastieri}, {Baughman}, {Bechtol}, {B{\'e}d{\'e}r{\`e}de}, {Bellardi}, {Bellazzini}, {Berenji}, {Bignami}, {Bisello}, {Bissaldi}, {Blandford}, {Bloom}, {Bogart}, {Bonamente}, {Bonnell}, {Borgland}, {Bouvier}, {Bregeon}, {Brez}, {Brigida}, {Bruel}, {Burnett}, {Busetto}, {Caliandro}, {Cameron}, {Caraveo}, {Carius}, {Carlson}, {Casandjian}, {Cavazzuti}, {Ceccanti}, {Cecchi}, {Charles}, {Chekhtman}, {Cheung}, {Chiang}, {Chipaux}, {Cillis}, {Ciprini}, {Claus}, {Cohen-Tanugi}, {Condamoor}, {Conrad}, {Corbet}, {Corucci}, {Costamante}, {Cutini}, {Davis}, {Decotigny}, {DeKlotz}, {Dermer}, {de Angelis}, {Digel}, {do Couto e Silva}, {Drell}, {Dubois}, {Dumora}, {Edmonds}, {Fabiani}, {Farnier}, {Favuzzi}, {Flath}, {Fleury}, {Focke}, {Funk}, {Fusco}, {Gargano}, {Gasparrini}, {Gehrels}, {Gentit}, {Germani}, {Giebels}, {Giglietto}, {Giommi}, {Giordano}, {Glanzman},
  {Godfrey}, {Grenier}, {Grondin}, {Grove}, {Guillemot}, {Guiriec}, {Haller}, {Harding}, {Hart}, {Hays}, {Healey}, {Hirayama}, {Hjalmarsdotter}, {Horn}, {Hughes}, {J{\'o}hannesson}, {Johansson}, {Johnson}, {Johnson}, {Johnson}, {Johnson}, {Kamae}, {Katagiri}, {Kataoka}, {Kavelaars}, {Kawai}, {Kelly}, {Kerr}, {Klamra}, {Kn{\"o}dlseder}, {Kocian}, {Komin}, {Kuehn}, {Kuss}, {Landriu}, {Latronico}, {Lee}, {Lee}, {Lemoine-Goumard}, {Lionetto}, {Longo}, {Loparco}, {Lott}, {Lovellette}, {Lubrano}, {Madejski}, {Makeev}, {Marangelli}, {Massai}, {Mazziotta}, {McEnery}, {Menon}, {Meurer}, {Michelson}, {Minuti}, {Mirizzi}, {Mitthumsiri}, {Mizuno}, {Moiseev}, {Monte}, {Monzani}, {Moretti}, {Morselli}, {Moskalenko}, {Murgia}, {Nakamori}, {Nishino}, {Nolan}, {Norris}, {Nuss}, {Ohno}, {Ohsugi}, {Omodei}, {Orlando}, {Ormes}, {Paccagnella}, {Paneque}, {Panetta}, {Parent}, {Pearce}, {Pepe}, {Perazzo}, {Pesce-Rollins}, {Picozza}, {Pieri}, {Pinchera}, {Piron}, {Porter}, {Poupard}, {Rain{\`o}}, {Rando}, {Rapposelli}, {Razzano},
  {Reimer}, {Reimer}, {Reposeur}, {Reyes}, {Ritz}, {Rochester}, {Rodriguez}, {Romani}, {Roth}, {Russell}, {Ryde}, {Sabatini}, {Sadrozinski}, {Sanchez}, {Sander}, {Sapozhnikov}, {Parkinson}, {Scargle}, {Schalk}, {Scolieri}, {Sgr{\`o}}, {Share}, {Shaw}, {Shimokawabe}, {Shrader}, {Sierpowska-Bartosik}, {Siskind}, {Smith}, {Smith}, {Spandre}, {Spinelli}, {Starck}, {Stephens}, {Strickman}, {Strong}, {Suson}, {Tajima}, {Takahashi}, {Takahashi}, {Tanaka}, {Tenze}, {Tether}, {Thayer}, {Thayer}, {Thompson}, {Tibaldo}, {Tibolla}, {Torres}, {Tosti}, {Tramacere}, {Turri}, {Usher}, {Vilchez}, {Vitale}, {Wang}, {Watters}, {Winer}, {Wood}, {Ylinen}, \& {Ziegler}}]{2009ApJ...697.1071A}
{Atwood}, W.~B., {Abdo}, A.~A., {Ackermann}, M., {et~al.} 2009, \apj, 697, 1071

\bibitem[{{Bamba} {et~al.}(2008){Bamba}, {Fukazawa}, {Hiraga}, {Hughes}, {Katagiri}, {Kokubun}, {Koyama}, {Miyata}, {Mizuno}, {Mori}, {Nakajima}, {Ozaki}, {Petre}, {Takahashi}, {Takahashi}, {Tanaka}, {Terada}, {Uchiyama}, {Watanabe}, \& {Yamaguchi}}]{2008PASJ...60S.153B}
{Bamba}, A., {Fukazawa}, Y., {Hiraga}, J.~S., {et~al.} 2008, \pasj, 60, S153

\bibitem[{Bruel {et~al.}(2018)Bruel, Burnett, Digel, Johannesson, Omodei, \& Wood}]{bruel2018fermilatimprovedpass8event}
Bruel, P., Burnett, T.~H., Digel, S.~W., {et~al.} 2018, Fermi-LAT improved Pass~8 event selection

\bibitem[{Condon {et~al.}(2017)Condon, Lemoine-Goumard, Acero, \& Katagiri}]{Condon_2017}
Condon, B., Lemoine-Goumard, M., Acero, F., \& Katagiri, H. 2017, The Astrophysical Journal, 851, 100

\bibitem[{Cristofari \& Blasi(2019)}]{Cristofari:2019pqh}
Cristofari, P. \& Blasi, P. 2019, Mon. Not. Roy. Astron. Soc., 489, 108

\bibitem[{{Giuffrida} {et~al.}(2022){Giuffrida}, {Miceli}, {Caprioli}, {Decourchelle}, {Vink}, {Orlando}, {Bocchino}, {Greco}, \& {Peres}}]{2022NatCo..13.5098G}
{Giuffrida}, R., {Miceli}, M., {Caprioli}, D., {et~al.} 2022, Nature Communications, 13, 5098

\bibitem[{{Hurley-Walker} {et~al.}(2017){Hurley-Walker}, {Callingham}, {Hancock}, {Franzen}, {Hindson}, {Kapi{\'n}ska}, {Morgan}, {Offringa}, {Wayth}, {Wu}, {Zheng}, {Murphy}, {Bell}, {Dwarakanath}, {For}, {Gaensler}, {Johnston-Hollitt}, {Lenc}, {Procopio}, {Staveley-Smith}, {Ekers}, {Bowman}, {Briggs}, {Cappallo}, {Deshpande}, {Greenhill}, {Hazelton}, {Kaplan}, {Lonsdale}, {McWhirter}, {Mitchell}, {Morales}, {Morgan}, {Oberoi}, {Ord}, {Prabu}, {Shankar}, {Srivani}, {Subrahmanyan}, {Tingay}, {Webster}, {Williams}, \& {Williams}}]{2017MNRAS.464.1146H}
{Hurley-Walker}, N., {Callingham}, J.~R., {Hancock}, P.~J., {et~al.} 2017, \mnras, 464, 1146

\bibitem[{{Kafexhiu} {et~al.}(2014){Kafexhiu}, {Aharonian}, {Taylor}, \& {Vila}}]{2014PhRvD..90l3014K}
{Kafexhiu}, E., {Aharonian}, F., {Taylor}, A.~M., \& {Vila}, G.~S. 2014, \prd, 90, 123014

\bibitem[{Katsuda {et~al.}(2009)Katsuda, Petre, Long, Reynolds, Winkler, Mori, \& Tsunemi}]{Katsuda2009}
Katsuda, S., Petre, R., Long, K.~S., {et~al.} 2009, The Astrophysical Journal, 692, L105

\bibitem[{{Koyama} {et~al.}(1995){Koyama}, {Petre}, {Gotthelf}, {Hwang}, {Matsuura}, {Ozaki}, \& {Holt}}]{1995Natur.378..255K}
{Koyama}, K., {Petre}, R., {Gotthelf}, E.~V., {et~al.} 1995, \nat, 378, 255

\bibitem[{Li {et~al.}(2018)Li, Ballet, Miceli, Zhou, Vink, Chen, Acero, Decourchelle, \& Bregman}]{Li_2018}
Li, J.-T., Ballet, J., Miceli, M., {et~al.} 2018, The Astrophysical Journal, 864, 85

\bibitem[{{Long} {et~al.}(2003){Long}, {Reynolds}, {Raymond}, {Winkler}, {Dyer}, \& {Petre}}]{2003ApJ...586.1162L}
{Long}, K.~S., {Reynolds}, S.~P., {Raymond}, J.~C., {et~al.} 2003, \apj, 586, 1162

\bibitem[{{Miceli} {et~al.}(2014){Miceli}, {Acero}, {Dubner}, {Decourchelle}, {Orlando}, \& {Bocchino}}]{2014ApJ...782L..33M}
{Miceli}, M., {Acero}, F., {Dubner}, G., {et~al.} 2014, \apjl, 782, L33

\bibitem[{{Miceli} {et~al.}(2012){Miceli}, {Bocchino}, {Decourchelle}, {Maurin}, {Vink}, {Orlando}, {Reale}, \& {Broersen}}]{2012A&A...546A..66M}
{Miceli}, M., {Bocchino}, F., {Decourchelle}, A., {et~al.} 2012, \aap, 546, A66

\bibitem[{{Miceli} {et~al.}(2016){Miceli}, {Orlando}, {Pereira}, {Acero}, {Katsuda}, {Decourchelle}, {Winkler}, {Bonito}, {Reale}, {Peres}, {Li}, \& {Dubner}}]{Miceli:2016sml}
{Miceli}, M., {Orlando}, S., {Pereira}, V., {et~al.} 2016, \aap, 593, A26

\bibitem[{{Sano} {et~al.}(2022){Sano}, {Yamaguchi}, {Aruga}, {Fukui}, {Tachihara}, {Filipovi{\'c}}, \& {Rowell}}]{2022ApJ...933..157S}
{Sano}, H., {Yamaguchi}, H., {Aruga}, M., {et~al.} 2022, \apj, 933, 157

\bibitem[{{Winkler} {et~al.}(2003){Winkler}, {Gupta}, \& {Long}}]{2003ApJ...585..324W}
{Winkler}, P.~F., {Gupta}, G., \& {Long}, K.~S. 2003, \apj, 585, 324

\bibitem[{{Winkler} {et~al.}(2014){Winkler}, {Williams}, {Reynolds}, {Petre}, {Long}, {Katsuda}, \& {Hwang}}]{2014ApJ...781...65W}
{Winkler}, P.~F., {Williams}, B.~J., {Reynolds}, S.~P., {et~al.} 2014, \apj, 781, 65

\bibitem[{{Wood} {et~al.}(2017){Wood}, {Caputo}, {Charles}, {Di Mauro}, {Magill}, {Perkins}, \& {Fermi-LAT Collaboration}}]{2017ICRC...35..824W}
{Wood}, M., {Caputo}, R., {Charles}, E., {et~al.} 2017, in International Cosmic Ray Conference, Vol. 301, 35th International Cosmic Ray Conference (ICRC2017), 824

\bibitem[{Xing {et~al.}(2016)Xing, Wang, Zhang, \& Chen}]{Xing2016}
Xing, Y., Wang, Z., Zhang, X., \& Chen, Y. 2016, The Astrophysical Journal, 823, 44

\bibitem[{{Xing} {et~al.}(2019){Xing}, {Wang}, {Zhang}, \& {Chen}}]{2019PASJ...71...77X}
{Xing}, Y., {Wang}, Z., {Zhang}, X., \& {Chen}, Y. 2019, \pasj, 71, 77

\bibitem[{{Zabalza}(2015)}]{2015ICRC...34..922Z}
{Zabalza}, V. 2015, in International Cosmic Ray Conference, Vol.~34, 34th International Cosmic Ray Conference (ICRC2015), 922

\end{thebibliography}
\bibliographystyle{aa}

\appendix
\section{\emph{Fermi}-LAT angular resolution}
\label{sec:ap}
\begin{figure}[t!]

\includegraphics[width=0.5\textwidth]{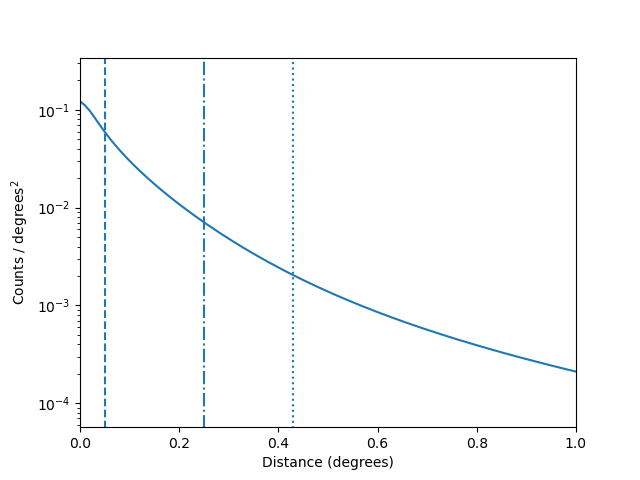}

\caption{\emph{Fermi}-LAT radial distribution of photons from a point source simulated with a spectral index of 2.0 between 1 GeV and 1 TeV, presented with a solid line. The dashed, dotted-dashed and dotted lines indicate the Half-Width at Half Maximum, the 50\% and 68\% containment radii, respectively.}
\label{fig:ang}
\end{figure}

As already discussed in the main text, the angular resolution of the LAT depends strongly on the gamma-ray's energy. Above 1 GeV, the point spread function (PSF) improves drastically, with a 68\% containment radius smaller than $0.2^{\circ}$ above 10 GeV. To better understand how this performance is characterized in our energy interval of interest (1 GeV—1 TeV for the morphological analysis in Section~\ref{sec:morpho}), we simulate a point source with a spectral index of 2.0 as a compromise between the hard component in the NE of the remnant and the softer component in the SW. Figure~\ref{fig:ang} shows the distribution of photons from this point source and the associated performances between 1 GeV and 1 TeV: the Half-Width at Half Maximum (HWHM) of $0.05^{\circ}$, which contains 13\% of photons in our case, the R50 (Half-Energy Width) of $0.25^{\circ}$ and the R68 (68\% containment radius) of $0.43^{\circ}$. The LAT point spread function is quite peaked but still contains large tails of photons as can be seen in this Figure. 

\end{document}